\documentclass[particles,article,accept,moreauthors,pdftex]{Definitions/mdpi} 
\firstpage{1} 
\makeatletter 
\DeclareUnicodeCharacter{2009}{\,} 
\makeatother
\pubvolume{xx}
\issuenum{1}
\articlenumber{5}
\pubyear{2020}
\copyrightyear{2020}
\history{Received: date; Accepted: date; Published: date}
\updates{yes} 

\Title{Neutron 
	Star Cooling Within the Equation of State With Induced Surface Tension}

\Author{Stefanos Tsiopelas and Violetta Sagun *
{$^{,\dagger}$}
\orcidB{}}

\AuthorNames{Stefanos Tsiopelas, Violetta Sagun}

\address[1]{%
CFisUC, Department of Physics, University of Coimbra, 3004-516 Coimbra, Portugal;}

\corres{\hangafter=1 \hangindent=1.05em \hspace{-0.82em} Correspondence: violetta.sagun@uc.pt}

\firstnote{\hangafter=1 \hangindent=1.05em \hspace{-0.82em} These authors contributed equally to this work.}

\abstract{We study the thermal evolution of neutron stars described within the equation of state with induced surface tension (IST) that reproduces properties of normal nuclear matter, fulfills the proton flow constraint, provides a high-quality description of hadron multiplicities created during the nuclear-nuclear collision experiments, and it is equally compatible with the constraints from astrophysical observations and the GW170817 event. The model features strong direct Urca processes for the stars above $1.91~M_{\odot}$. The IST equation of state shows very good agreement with the available cooling data, even without introducing nuclear pairing. We also analysed the effect of the singlet proton/neutron and triplet  neutron pairing  on the cooling of neutron stars of different mass. We show that the description of the compact object in the center of the Cassiopeia A does not necessarily require an inclusion of neutron superfluidity and/or proton superconductivity. Our results indicate that data of Cassiopeia A can be adequately well reproduced by a $1.66~M_{\odot}$ star with an atmosphere of light elements. Moreover, the IST EoS reproduces each of the observational datasets for the surface temperature of Cassiopeia A either by a rapidly cooling $\sim$$1.955~M_{\odot}$ star with paired and unpaired matter or by a $1.91~M_{\odot}$ star with the inclusion of neutron and proton pairings in the singlet channel.
}


\begin{document}

\section{Introduction}
\label{Intro}

Born out of supernova explosions, neutron stars (NSs) are considered to start their life having very high internal temperatures and cool down through a combination of thermal radiation from their surface and neutrino emission from their interior. From the first day of their lives, when the temperature of their interior has already dropped from $\sim$$10^{11}$ $K$ to $ \sim$$10^{9-10}$ $K$ making it transparent to neutrinos, up until the first million years of their existence, thermal energy is mainly carried away in the form of neutrino radiation. During this time, measurements of surface temperature and luminosity of the stars can provide significant information about the properties of matter in their depth, since the thermal evolution depends on factors, such as the internal composition and the thermodynamic properties of matter, are defined by the Equation of State (EoS), the chemical abundances of the envelope, and the type of pairing between the constituent particles. Therefore, it is necessary for any theoretical calculations aiming to describe the cooling process of NSs to consider the various combinations between those factors. The set of cooling curves extracted from these simulations can then be tested against the observable features of NSs in the X-ray part of the spectrum.

The particle composition of the core of NSs plays a prominent role in their cooling process, since it is the decisive factor of whether the direct Urca (DU) process of neutron $\beta$-decay and its inverse process are allowed to occur in the interior of the star \cite{1991PhRvL..66.2701L}. These are the most efficient neutrino-emitting reactions that can happen in hadronic matter and, when permitted, they lead to a rapid cooling of the NS, substantially different from the case they are forbidden \cite{Page2006,Potekhin2015}. For these reactions to be possible, the Fermi momenta of the participating particles must satisfy the kinematic restriction of triangle inequality. While taking charge neutrality and the relation between the Fermi momenta and the number density of each particle into account, this constraint leads to the requirement that the proton fraction should be higher than $\sim$$11\% $ of the total baryon density \cite{Page2006}.

The composition and thermodynamic properties of compact stars were described by the EoS formulated within the framework of induced surface tension (IST). Originally, this approach was developed in order to correctly account for the short-range repulsion between the particles characterized by the different hard-core radii, which are known to be principal elements of hadron EoS at high temperatures typical for collisions of heavy ions \cite{2006NuPhA.772..167A}. One of the main advantages of the IST EoS compared to other approaches considering the hard-core repulsion is its ability to correctly account for the excluded volume of particles reflected in correct values of first four virial coefficients of hard spheres \cite{2018NuPhA.970..133B}. Thus,~the IST EoS goes well beyond the usual van der Waals approximation, which increases the range of its causality and applicability. This allowed the present model to describe the nuclear matter phase diagram at low temperatures and intermediate densities, the region of the nuclear liquid-gas phase transition and its critical end-point \cite{violetta1}. In principle, the present model, in comparison to the Skyrme and Relativistic Mean Field models, can be extended for an arbitrarily large number of different particle species, as was done in Ref. \cite{2018NuPhA.970..133B}. This feature of the present approach also allowed it to successfully describe hadron yields that were measured in heavy-ion collision experiments, i.e AGS (Alternating Gradient Synchrotron, Brookhaven National Laboratory), SPS (Super Proton Synchrotron, CERN), RHIC (Relativistic Heavy Ion Collider, Brookhaven National Laboratory), and LHC (Large Hadron Collider, CERN) \cite{violetta2, Sagun2017a}. Extensive analysis of nuclear-nuclear collision experimental data with the multicomponent IST EoS helped to constraint the repulsive interaction between the particles, described by the hard-core radii. The successful application of the IST approach in a wide range of temperatures and densities demonstrated its potential towards formulation of a unified EoS of hadron matter. Further development of this model led to incorporation of the long-range interaction within the mean-field framework \cite{Ivanytskyi-Bugaev2018}, which allowed for the IST EoS to fulfill the flow constraint of Danielewicz et al. \cite{2002Sci...298.1592D}. Recently, the present model was applied to the modeling of nuclear matter at very high density and vanishing temperature existing inside the compact astrophysical objects  \cite{LS1,LS2,Sagun-Lopes22019}. The present study of the thermal evolution of NSs is the next step towards constraining the IST EoS and developing a unified hadron EoS.

For a realistic description of the outer layers of the NS the IST EoS is supplemented by the Haensel-Zdunik (HZ) EoS for the outer crust and the Negele--Vautherin (NV) EoS for the inner crust~\cite{Haensel1990,Negele1973}.

An additional important factor worth considering is the superfluidity (superconductivity) of neutrons (protons) in the NS interior \cite{1976ApJ...205..541F}, since they are able to alter the thermal evolution of the star~\cite{Yakovlev-Kaminker2001}. Despite suppressing the rates of neutrino emission from the rest of the processes, pairing between protons or neutrons introduces a new $\nu_{e}-$emitting mechanism, due to the permanent formation and breaking of Cooper pairs, known as PBF \cite{Potekhin2015}. This neutrino emitting process is activated in each baryon species as soon as the temperature reaches the respective critical temperatures $ T_{c,n},$ $T_{c,p}$. Nevertheless, the details, such as the onset T, the form of the gap parameter, and the peak of the PBF emissivity, are strongly sensitive to the gap model that was adopted to describe the matter in this supercritical state. As a result, applying different models to study the NS cooling provides also the possibility of gaining a further insight on the behavior of paired matter.

One of the most intriguing and debatable object that serves as a test ground for many models of nucleon pairing and alternative approaches is Cassiopeia A (Cas A). This compact source being a very young (at an age of $\approx 340$ yr) NS demonstrates a rapid cooling across a decade of observations. Thus, since the early observations by Heinke\&Ho (2010), the reported surface temperature change is decreasing from $3.6\pm0.6\%$ \cite{2010ApJ...719L.167H} to $3.5\pm0.4\%$ \cite{2013ApJ...777...22E}, recently refined to $<$$2.4\%$ ($<$$3.3\%$) for constant (varying) hydrogen absorption \cite{2018ApJ...864..135P}, and $2.1\pm0.2\%$ ($2.7\pm0.3\%$) for constant (varying) hydrogen absorption  \cite{2019MNRAS.484..974W}.
Moreover, as it was reported by Ho \& Heinke (2009) \cite{Ho2009}, the Cas A X-ray spectra suggest a carbon composition of the atmosphere and a mass in a range between $1.4~M_{\odot}$ and $2.4~M_{\odot}$ for the NS in question. An interpretation of such a rapid cooling of the Cas A includes onset of superfluidity (superconductivity) in the core of the star \cite{2011MNRAS.411.1977Y, 2011MNRAS.412L.108S, 2011PhRvL.106h1101P,Taranto2016}, nuclear medium cooling scenario~\mbox{\cite{2012PhRvC..85b2802B, 2013PhRvC..88f5805B}}, phase transition to quark-gluon plasma \cite{2013A&A...555L..10S, 2013ApJ...765....1N}, decay of magnetic field  \cite{2014A&A...561L...5B}, heating by r-mode oscillations~\cite{2011ApJ...735L..29Y}, etc. Unfortunately, for the time being, available observational data do not provide us with a possibility to differentiate between these scenarios.

Based on the IST EoS, in this work we study the thermal evolution of NSs, aiming to describe the available cooling data. Our investigation is focused on the implications of employing various gap models in order to describe neutron superfluidity and proton superconductivity. In addition, we examine the effect of envelope composition on the cooling process of compact stars. Modeling the thermal evolution of NSs was performed by using the {\it NSCool} code \cite{Page2006,PageReddy2006}. 

The paper is organized, as follows: in Section \ref{EOS}, we present a brief description of the model. In Section \ref{cool}, we discuss the neutrino emission mechanisms taking place in a NS as well as the pairing models used in the simulations. Section \ref{res} is dedicated to the results of the calculations, while Section \ref{Concl} contains the summary and the conclusions.

\section{Equation of State}
\label{EOS}

In the present formulation, the IST EoS includes neutrons, protons, and electrons, assuming that an effect of heavy baryons and mesons, which can appear at high densities, is absorbed in the mean-field potential. Thus, the model accounts for the strong short-range repulsion and relatively weak long-range attraction between nucleons, while electrons are treated as an ideal Fermi gas. The former part of the nucleon-nucleon interaction is modelled with the hard-core radius, similarly to the famous Van der Waals EoS. The hard-core repulsion of nucleons leads to an appearance of an excluded volume. However,~contrary to the Van der Waals approximation, the IST EoS instead of the constant excluded volume has a density-dependent one. The key element of the model is the IST coefficient. It accounts not only for the correct values of four virial coefficients of hard spheres, but it also extends the causality range of the model to the density range typical for the NSs \cite{LS2, violetta1}. {Moreover, the long-range attraction and asymmetry between neutrons and protons are accounted via the mean-field potentials, the parameters of which were fitted to the properties of matter at saturation density, and, therefore, effectively accounts for many particles interaction of different species (for details see \cite{Ivanytskyi-Bugaev2018}).} More detailed information regarding the model and its application to the NSs can be found in Refs. \cite{LS2, Sagun-Lopes22019, violetta1}.

\begin{figure}[H]
\centering
\includegraphics[scale=0.55]{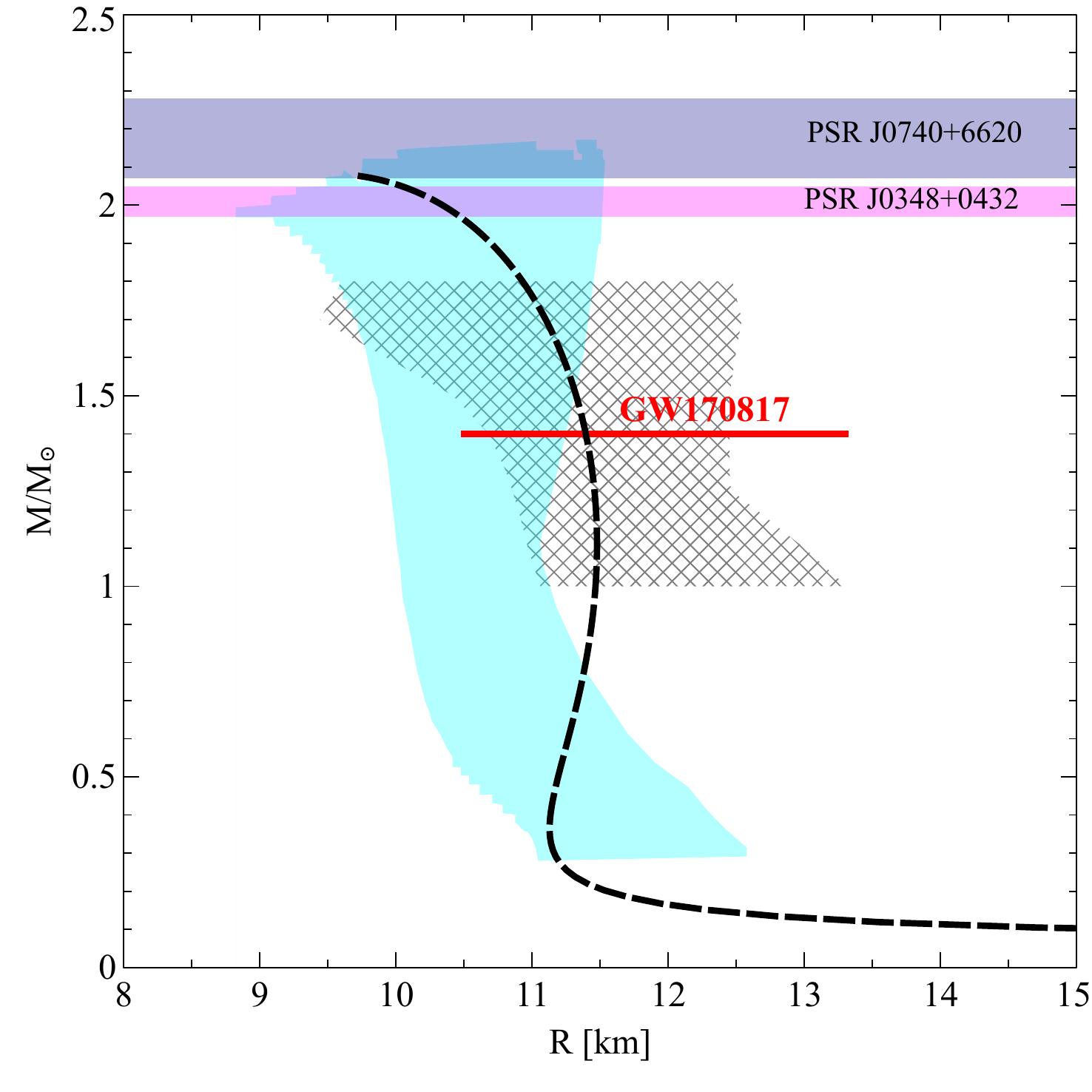} \\
\caption{Mass-radius relation for non-rotating neutron stars (NSs) calculated for set B of the induced surface tension Equation of State (IST EoS) \cite{Sagun2020}. Horizontal bands correspond to the two most massive NSs, e.g., PSR J0348+0432 \cite{2013Sci...340..448A} (magenta band) and PSR J0740+6620 \cite{2020NatAs...4...72C} (blue band). The shaded grey area represents the M-R constraint taken from \mbox{Refs. \cite{2010ApJ...722...33S, 2013ApJ...765L...5S}}, while the constraint depicted as a cyan area was taken from Ref. \cite{2016ARA&A..54..401O}. The red line represents the allowed range of NS radius, according to GW170817 event.}                                                                                                                                                                                                                                                                                                                                                                                                                                                                                                                                                                                                                                                                                                                                                                                                                                                                                                                                                                                                                                                                                                                        
\label{fig:1} 	
\end{figure}

We adopt the softer parameterisation of the IST EoS that corresponds to the set B of Ref. \cite{Sagun2020} \mbox{(for simplicity} we keep referring to this parameterisation of the model as set B). The set B was chosen due to the better agreement with the astrophysical constraints and results coming from the GW170817~\cite{LIGO}. It gives the values of the symmetry energy $E_{sym}= 30.0$ MeV, symmetry energy slope $L= 93.2$ MeV and nuclear incompressibility factor $K_{0}= 201.0$ MeV at normal nuclear density that is in full agreement with the present nuclear matter results \cite{LS2, Sagun-Lopes22019}.
As shown in Figure \ref{fig:1}, the maximum mass $M_{max}=2.08 M_{\odot}$ is consistent with the recent measurements of the most massive NSs, i.e., PSR J0348+0432 \cite{2013Sci...340..448A} and PSR J0740+6620 \cite{2020NatAs...4...72C}.

\begin{figure}[H]
\centering
\includegraphics[scale=0.55]{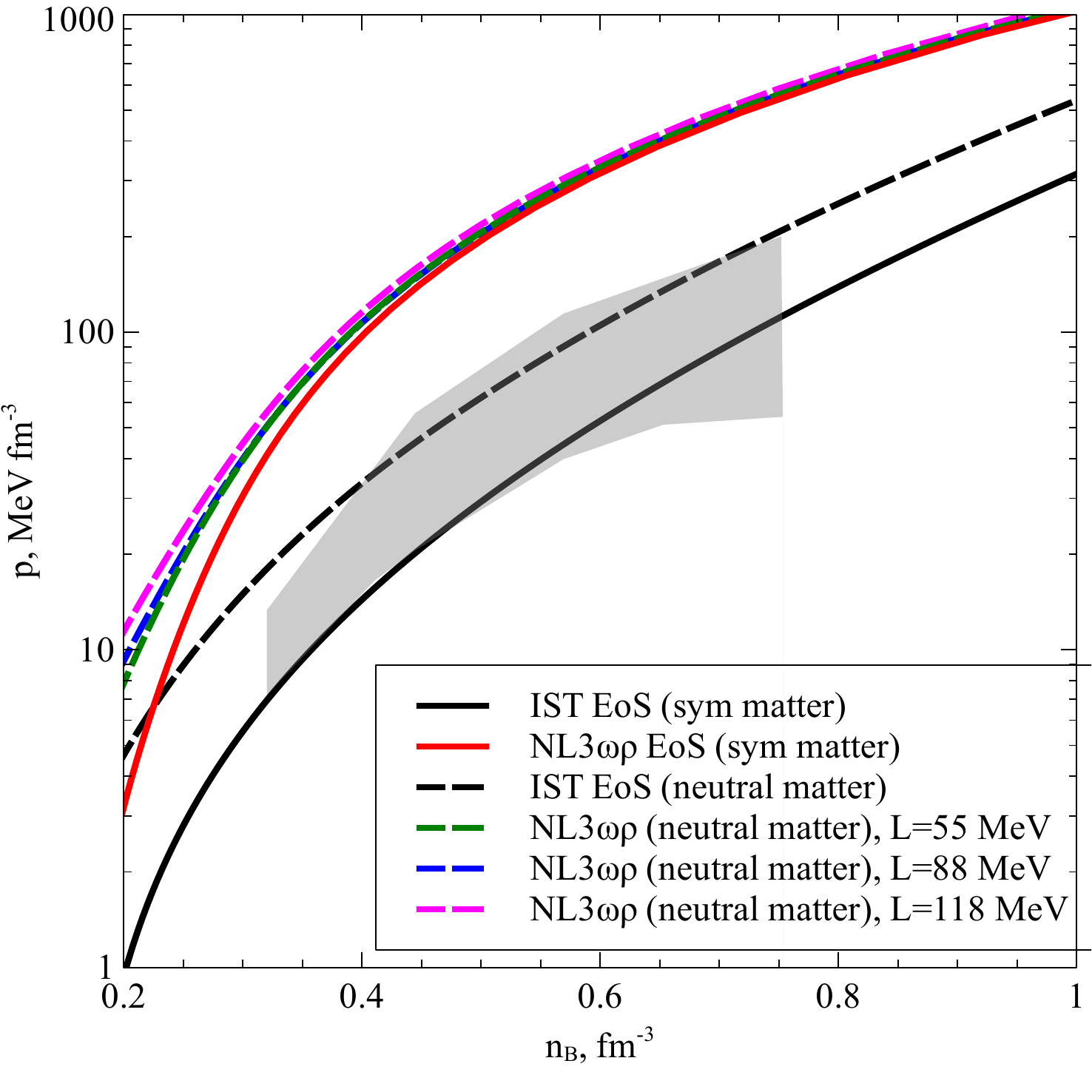} \\
\caption{Pressure as a function of baryonic density for symmetric and $\beta$-equilibrated nuclear matter for the IST EoS (solid and dashed black curves, correspondingly) and the $NL3\omega\rho$ EoS. The solid red curve corresponds to the $NL3\omega\rho$ EoS for symmetric matters (or NL3, for simplicity). The  $NL3\omega\rho$ EoS for the $\beta$-equilibrated nuclear matter is shown for L = 55 MeV (dashed green curve), L = 88 MeV (dashed blue curve) and L=118 MeV (dashed magenta curve). The shaded gray area corresponds to the flow constraint for symmetric matter \cite{2002Sci...298.1592D}.}                                                                                                                                                                                                                                                                                                                                                                                                                                                                                                                                                                                                                                                                                                                                                                                                                                                                                                                                                                                                                                                                                                                        
\label{fig:1b} 	
\end{figure}

The relatively high slope of the symmetry energy of the present model requires a special comment. The analysis of correlation between this quantity and the radius of a $1.4~M_{\odot}$ NS performed within a wide set of Skyrme and Relativistic mean-field (RMF) models shows that $L>90$ MeV is unlikely to be consistent with $R_{1.4}$ below 12 km \cite{2016PhRvC..94e2801A, 2020PhRvC.102d5807L, 2013PhLB..727..276L, 2017RvMP...89a5007O}. However, as seen from Figure \ref{fig:1}, the present model yields $R_{1.4}=11.39$ km. In order to explain this apparent discrepancy, we have to mention that a stellar characteristic, such as the NS radius, is defined by the total stiffness of an EoS. The considered model contains an isospin-independent term and a term related to the nuclear symmetry energy. While high values of $L$ lead to stiff symmetry energy, this redundant stiffness can be moderated by the soft isospin-independent part of EoS, which is the case for the IST EoS. In order to demonstrate this, we compare EoS of symmetric nuclear matter calculated within the present model and the well known RMF model NL3 \cite{1997PhRvC..55..540L, 2003ApJ...593..463C}. As shown on Figure \ref{fig:1b}, the symmetric IST EoS is indeed not only significantly softer than the symmetric NL3 model, but it also allows for the present model to fit the proton flow constraint~\cite{2002Sci...298.1592D} (the shaded gray area on Figure \ref{fig:1b}). Further, we compare the electrically neutral IST EoS with the three parameterizations of RMF model $NL3\omega\rho$, which implies coupling of $\omega$ and $\rho$ mesons leading to different slopes of the symmetry energy \cite{2016PhRvC..94a5808P}. The values of $L=$ 55~MeV, 88~MeV and 118~MeV (green, blue and magenta dashed curves, respectively) label the $NL3\omega\rho$ models. It is seen from Figure \ref{fig:1b} that, despite its large $L$, the neutral IST EoS is actually rather soft, which makes it consistent with $R_{1.4}<$ 12~km.

\section{Cooling Processes}
\label{cool}

The thermal evolution of NSs can be divided in two stages. During the first one, which is known as neutrino cooling era, the $\nu_{e}$ emission generated from a plethora of emission mechanisms throughout the whole interior of the star dominates the cooling process \cite{Page2006,Potekhin2015}. On a timescale of a million years after the NS formation, when $T_{core}$ has dropped below $10^{8}\,K$, neutrino emission from the core is less efficient due to the strong temperature dependence, and photon emission from the surface overtakes as the leading heat loss mechanism. Beyond this point, the neutrino cooling era is over and the photon cooling era begins. This shift is marked by an abrupt decline of the total luminosity and surface temperature of the star, with more massive stars exhibiting steeper drops of those quantities than less massive ones, since photon luminosity follows the typical blackbody radiation law.

During the neutrino cooling era, the leading neutrino-generating process varies over the regions of the star and over time. Overall, the main factors regulating the neutrino emissivity of each process are density, temperature, and the existing degree of Cooper pairing between particles. For example, in the envelope of the star a pair annihilation is the most productive mechanism, while in the outer crust neutrino emission is dominated by plasmon decay until its temperature reaches a few times $10^{8}\, K$, replaced by electron-ion bremsstrahlung beyond that point \cite{Yakovlev-Kaminker2001, Potekhin2015}. The latter remains the most efficient energy loss mechanism throughout the inner crust as well, as long as the neutrons of this region are paired \cite{Kaminker2001}. Even in their supercritical state though, inner crust neutrons contribute to the total neutrino emission through the PBF process over a narrow range of densities and temperatures \cite{Potekhin2015}.

In the core of the star, apart from the bremsstrahlung processes between the free particles, the main neutrino-emitting processes that can occur are the DU process of the neutron $\beta$-decay and its~inverse:
\begin{align}
n  \rightarrow p + e + \bar{\nu}_{e} ,~~~~
p + e  \rightarrow  n + \nu_{e} \;.
\end{align}

The realization of these reactions depends on the proton fraction in the stellar interior, as mentioned above. For the IST EoS adopted in this work, these fast processes can only proceed in the core of stars with central densities $ n_{c} $ higher than $n_{DU}=0.862\, fm^{-3}$. For NSs with lower $ n_{c} $, where the Fermi momenta of the involved particles fail to satisfy the kinematic restriction of $p_{F,n} \leq p_{F,p} + p_{F,e} $, neutron $\beta$-decay and its inverse can still proceed with the help of a spectator proton or neutron that provides the extra $ p_{F,i} $ needed for the conservation of momentum. In this case, though, the efficiency of these so-called modified Urca (MU) processes in removing heat from the star is lower than that of the DU ones, resulting in an overall slower cooling of the star. Nevertheless, neutrino emission from MU processes surpasses the rest of the emissivities in the core, unless n and p are in a paired state, which, in turn, results in a further suppression of the neutrino emission rates. However, pairing partially compensates for the delay in the cooling of the NS core as it introduces PBF, an additional efficient channel for carrying the heat away in the form of neutrino-antineutrino pairs \cite{Potekhin2015}.

Cooper pairing of core protons and crust neutrons sets in a few years after the NS birth, while core neutrons pairing is viable at a later time. As regards the types of pairing, all of them follow the standard BCS pattern, with free neutrons of the inner crust and protons of the core undergoing singlet-state pairing $(^{1}S_{0})$, while neutrons of the core are expected to pair in the triplet-state $(^{3}P_{2})$~\mbox{\cite{Becker2009,Page2006,Bardeen1957}}. Although a complete, precise description of the effect for densities that are relevant to NS matter is still pending, it is widely applied for investigating its imprint on the thermal evolution of compact stars. The transition of neutrons (protons) to superfluid (superconducting) state also inflicts a suppression on their specific heat. Most studies agree that, once it switches on for a certain species, $ c_{\upsilon,i} $ is reduced by a factor $\mathcal{R}\sim e^{-\Delta/T}$, where $\Delta$ is the gap parameter, linked to the respective critical temperature $T_{c}$ via the standard BCS pairing relation $T_{c} \approx 0.57\Delta $ \cite{Becker2009,Page2006,Bardeen1957}. A further drop in temperature causes a further reduction of heat capacity to the point of being equal to that of leptons \cite{Potekhin2015}. Concerning the neutrino emissivity of processes that involve paired baryons, suppression is also induced, since particles in such a supercritical state have to overcome the energy gap, in order to interact with another particle. In the range of temperatures $T\ll T_{c}$, this behavior is numerically illustrated while using a yet another set of control functions $ \mathcal{R}_{\chi}\left(T/T_{c}\right)$, which differs between each particle species $\chi$\; \cite{Yakovlev-Kaminker2001}.

The onset temperatures $T_{c,n} (T_{c,p})$ of superfluidity (superconductivity), the associated $\Delta$, and the profile of PBF neutrino emissivity vary over the several models developed to describe the paired matter. Therefore, when including nucleon pairing for simulating the thermal evolution of NSs, the effect is subject to the gap models employed. Our choice for the simulations were the following models: SFB  \cite{Schwenk2003} for the  $^{1}S_{0}$ superfluidity of neutrons, T72  \cite{Takatsuka1972} and AO  \cite{Amundsen1985a} for their $^{3}P_{2}$ superfluidity, and AO \cite{Amundsen1985b}, CCDK \cite{Chen1993} for the proton $^{1}S_{0}$ channel in different combinations. The critical temperatures $ T_{c} $ were calculated according to the phenomenological formula that was suggested by Kaminker et al. and the parameters used by Ho et al. \cite{Kaminker2001,Ho2015}. It has to be to noted, though, that the vertex corrections were omitted in the study.

\begin{figure}[H]
\centering
\includegraphics[scale=0.55]{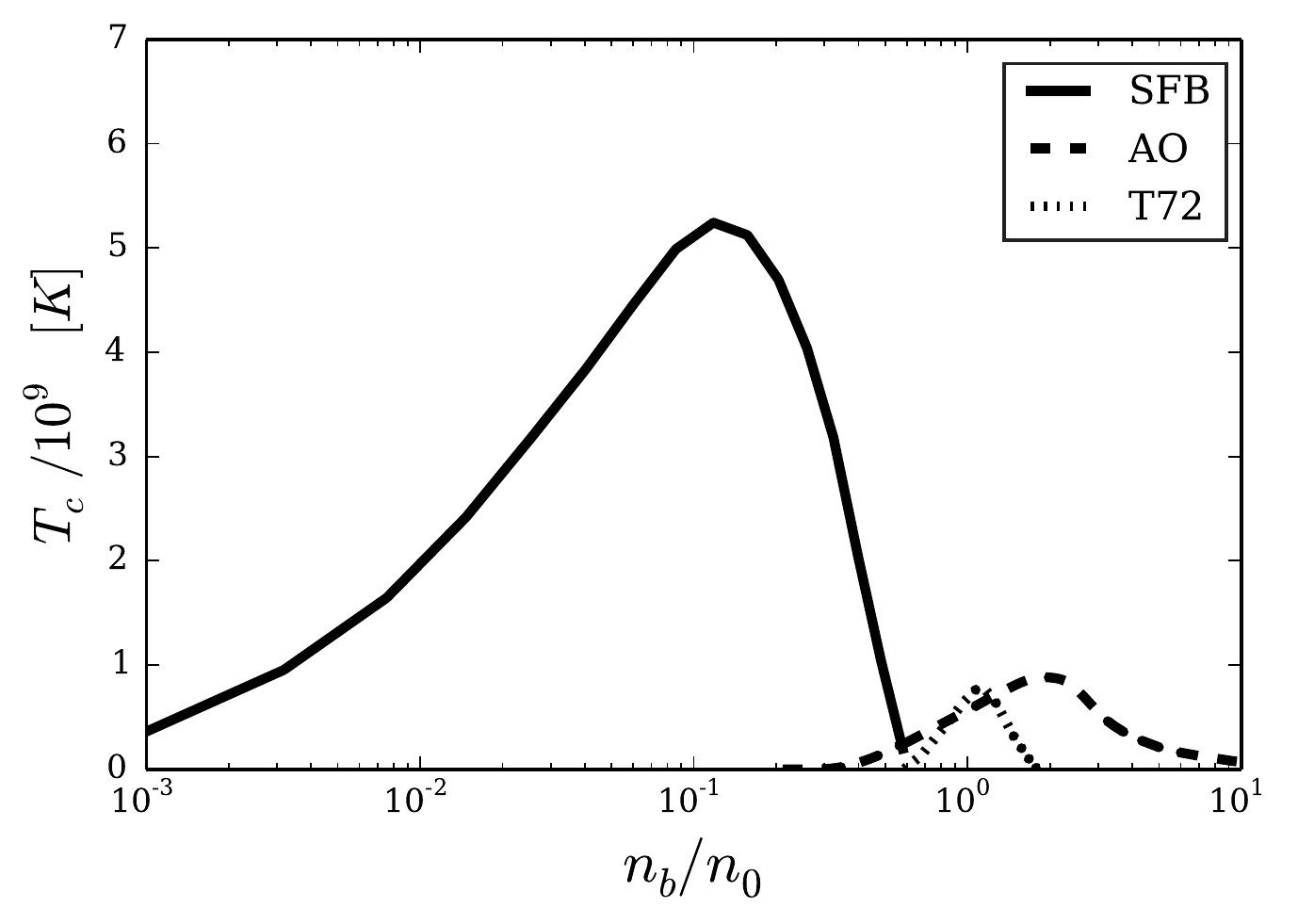}
\includegraphics[scale=0.55]{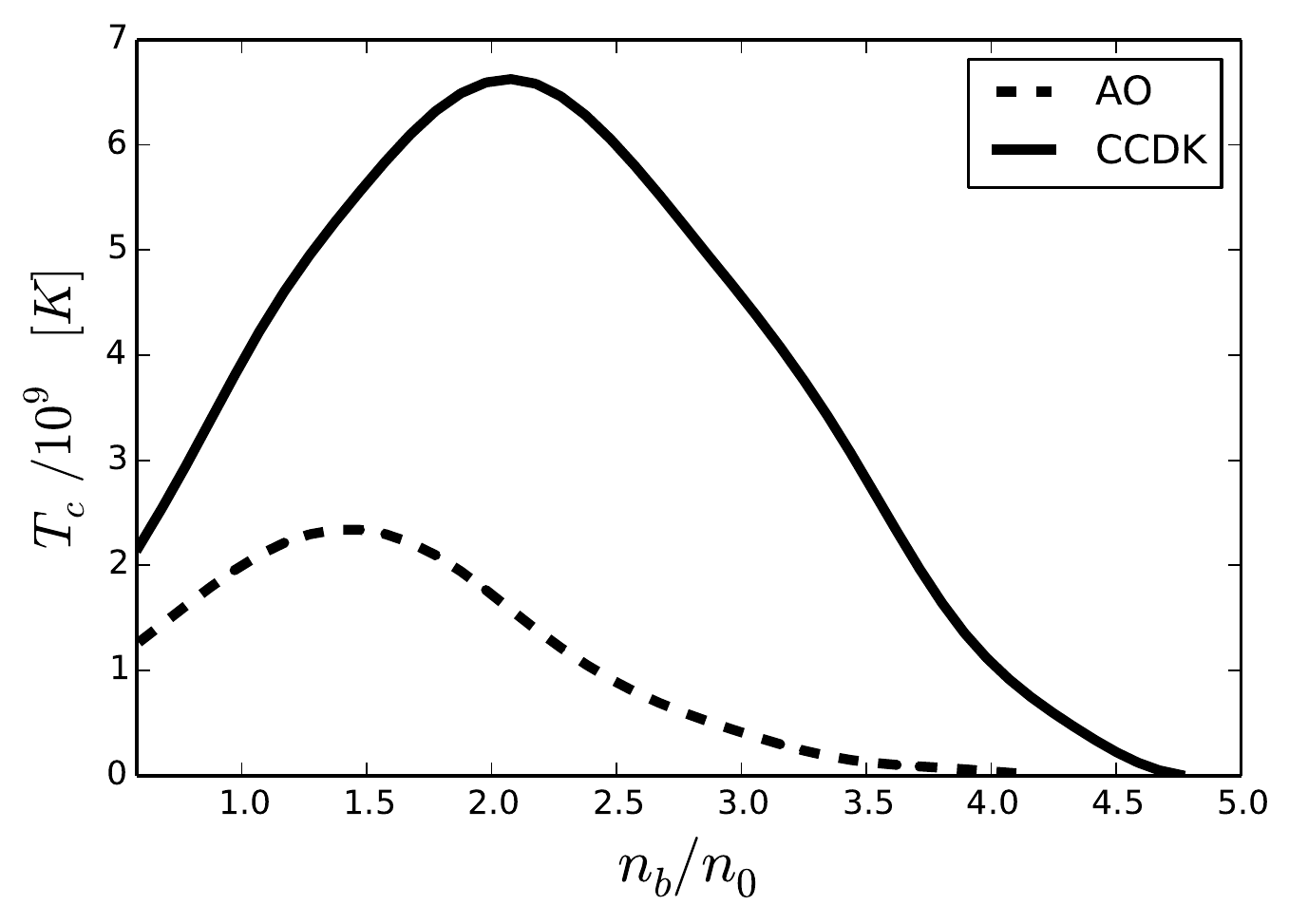}
\caption{Density dependence of the critical temperature for the considered singlet and triplet neutron gaps (left panel) and proton singlet pairing gaps (right panel).}                                                                                                                                                                                                                                                                                                                                                                                                                                                                                                                                                                                                                                                                                                                                                                                                                                                                                                                                                                                                                                                                                                                        
\label{fig:2} 	
\end{figure}

The relation between the critical temperature and the baryon density for all of the adopted gap models is shown on Figure \ref{fig:2}. According to the SFB model, $^{1}S_{0}$ pairing of crust neutrons appears first at densities around $ 0.15\,n_{0} $, once the crust temperature has dropped below $ \sim 5\cdot10^{9}\, K $. As the crust keeps cooling, the region of superfluid crust neutrons expands both outwards and towards the crust-core interface. Regarding the superconductivity of protons in the core, the CCDK model suggests that the proton pairing occurs initially at a temperature not greater than $ 7\cdot10^{9}\, K $ in the layer that corresponds to $\sim 2\,n_{0}$. On the contrary, the AO model proposes that the onset temperature is a few times lower, at $ 2\cdot10^{9}\, K $, while the associated baryon density only slightly lower. Finally, as claimed by the T72 model, the threshold for the onset of $^{3}P_{2}$ superfluidity of core neutrons is below $ 10^{9}\, K $ and pairing occurs in a narrow, symmetric region centered around the normal nuclear density $n_{0}$. The respective AO model, on the other hand, although it does not differ considerably in terms of $T_{c,n}$, implies that neutron pairing can occur throughout the whole region of the core.

The composition of the envelope is another deciding factor for the surface photon luminosity of the star, which is ultimately the quantity of observational importance. Namely, heavier elements tend to delay heat transport from the outer crust to the surface, since, in this case, the electron thermal conductivity is reduced \cite{Yakovlev1980}. The envelope acts as a blanket that, due to its poor thermal conductivity, creates a temperature gradient between the surface and interior of the star. The standard approach of thermal evolution codes is to set an envelope model as an outer boundary condition that links the temperature at the bottom of the envelope $(T_{b})$ to the temperature of the stellar surface $(T_{s})$. The fundamental assumptions behind this method are that the envelope has a thermal relaxation timescale that is much shorter than that of the crust, and that the neutrino emissivity in the envelope is negligible \cite{Page2006}. In this work, we used two distinct envelope models: one composed of heavy elements  and a hydrogen-rich one that contains the fraction of light elements $\eta=\Delta M/M= 10^{-7}$ \cite{Potekhin2003}. The presence of light elements (such as H) in the outer layers may suggest an accretion onto the NS after its formation, elements between He and C could occur as a result of diffusive nuclear burning on the surface of the star, while an existence of much heavier elements (up to Fe) give a hint of no accretion or the thermonuclear reactions after it \cite{2019MNRAS.484..974W}.

\section{Results}
\label{res}

\begin{figure}[H]
\centering
\includegraphics[scale=0.55]{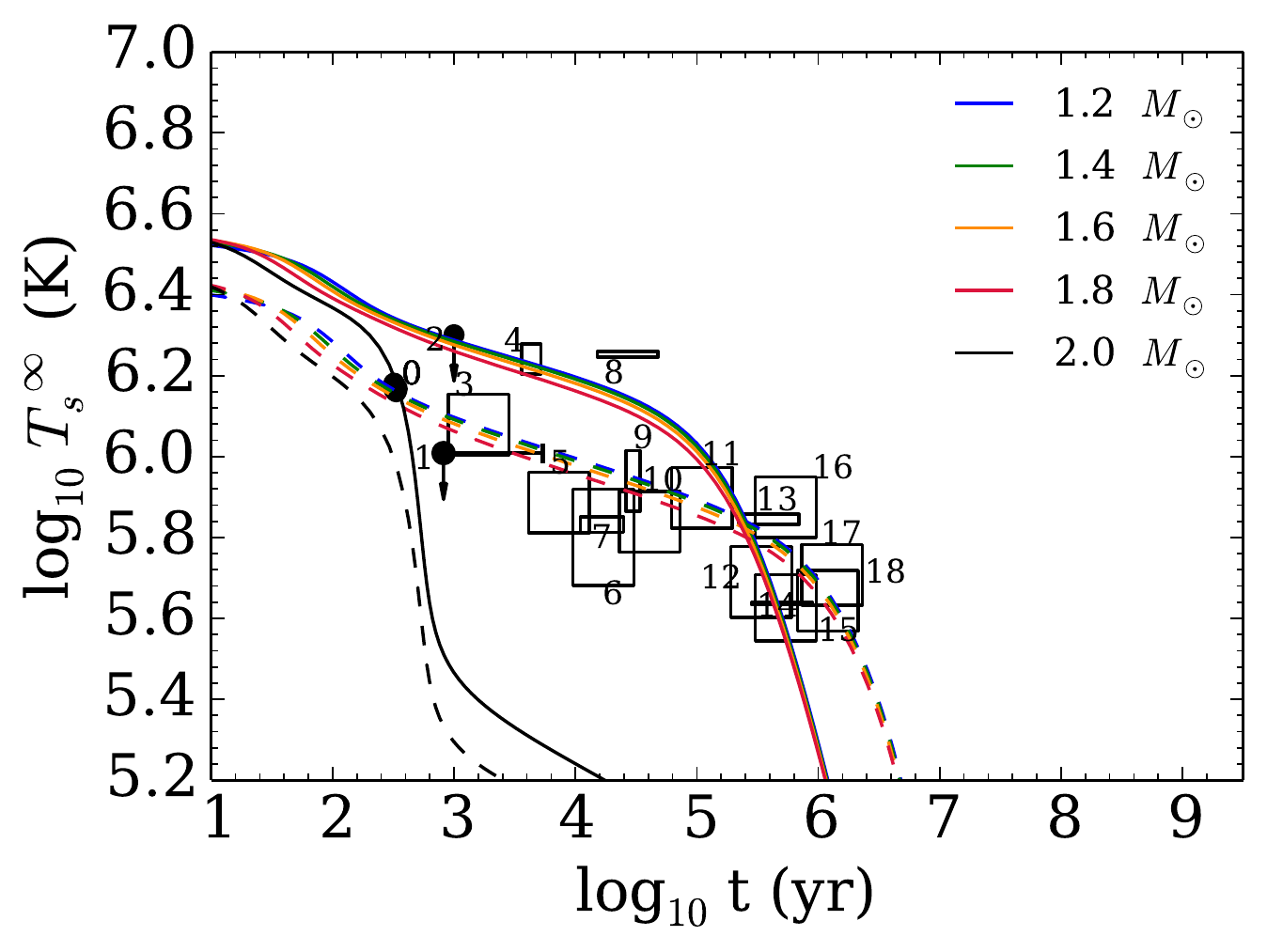} \\
\caption{Cooling curves for stars of different mass $M/M_{\odot}=1.2, 1.4, 1.6, 1.8, 2.0$ for the case of unpaired matter. $T^{\infty}_{S}$ denotes the surface temperature at infinity. The solid curves correspond to the light-element envelope ($\eta=10^{-7}$), while the dashed curves were obtained for the heavy-element envelope. The data points are taken from \cite{2015MNRAS.447.1598B}.
}                                                                                                                                                                                                                                                                                                                                                                                                                                                                                                                                                                                                                                                                                                                                                                                                                                                                                                                                                                                                                                                                                                                        
\label{fig:3} 	
\end{figure}

At the beginning, we focused on the thermal evolution of NSs without any sort of pairing between the nucleons. As you can see on Figure \ref{fig:3}, all of the cooling curves that are depicted in color exhibit a slow cooling due to the domination of the MU processes, since the DU ones are not kinematically allowed. Finally, at~central densities of over $n_{DU}=0.862\, fm^{-3}$ of beta-stable and charge neutral matter, which correspond to NSs with masses $M \geq 1.91\, M_{\odot}$, the DU processes are switched on in the core of the star, so that it undergoes enhanced cooling (see the black curves on Figure \ref{fig:3}). In order to model the uncertainties of the heat-blanketing effect of the envelope, we compare the thermal evolution of NSs with a non-accreted envelope containing heavy elements (dashed curves on Figure \ref{fig:3}) with the envelope containing light elements (solid curves on Figure \ref{fig:3}).
Remarkably, the obtained cooling curves for unpaired matter describe the experimental data very good. 

\begin{figure}[H]
\centering
\includegraphics[scale=0.55]{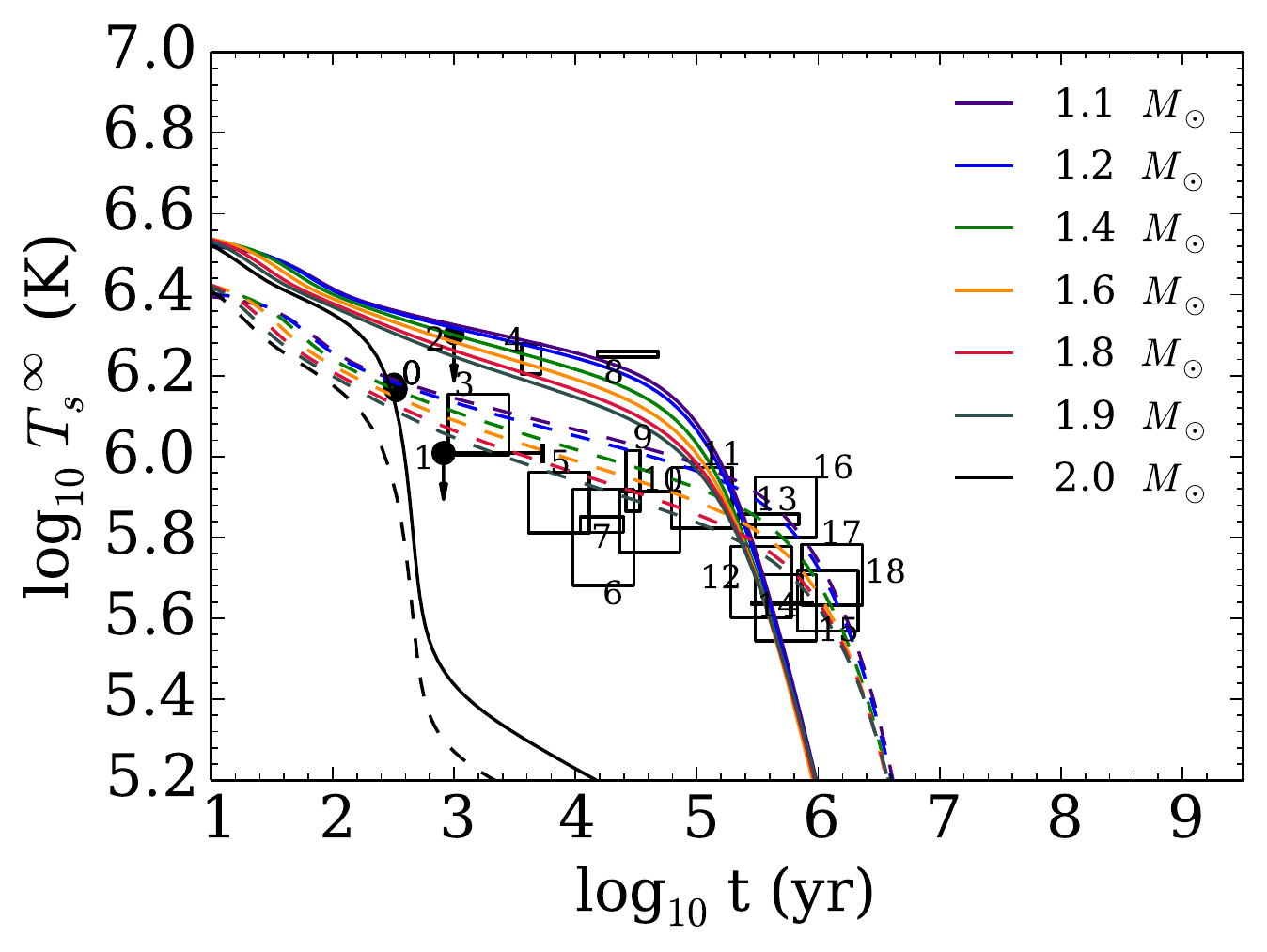}   
\includegraphics[scale=0.55]{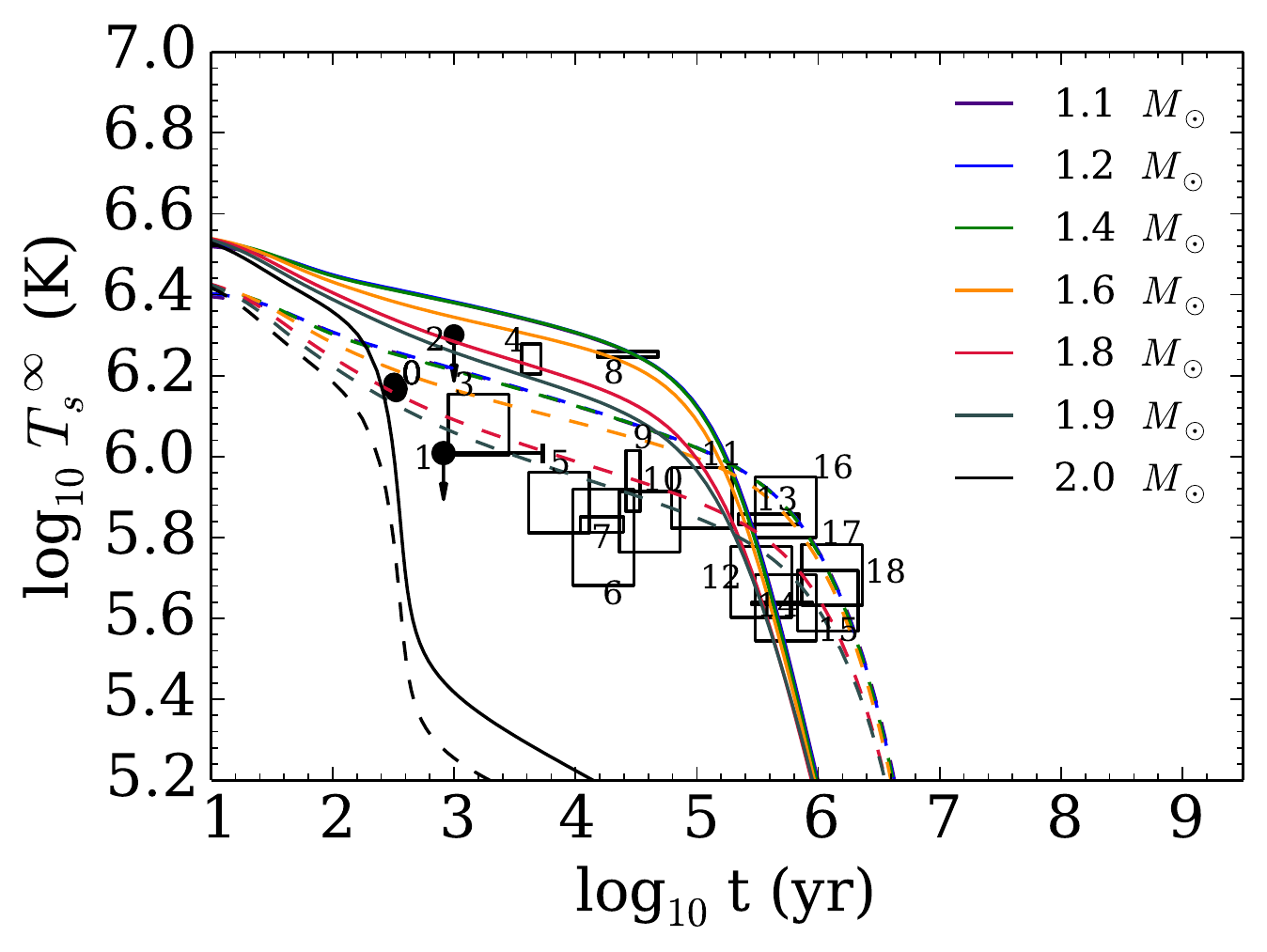} 
\caption{The same as Figure \ref{fig:3}, but considering the effect of neutron superfluidity in the $^{1}S_{0}$ channel via the SFB model \cite{Schwenk2003} and proton superconductivity in the $^{1}S_{0}$ channel with the AO model \cite{Amundsen1985a} \mbox{(left panel)} and CCDK model \cite{Chen1993} (right panel).}                                                                                                                                                                                                                                                                                                                                                                                                                                                                                                                                                                                                                                                                                                                                                                                                                                                                                                                                                                                                                                                                                                                        
\label{fig:4} 	
\end{figure}

Furthermore, we studied the effect of neutron superfluidity and proton superconductivity on the thermal evolution of NSs in two stages. First, we considered n$^{1}S_{0}$ superfluidity with the SFB model~\cite{Schwenk2003} together with a p$^{1}S_{0}$ superconductivity that was described by AO \cite{Amundsen1985b} and CCDK \cite{Chen1993} models \mbox{(see Figure \ref{fig:4})}. For both considered combinations of gaps, i.e. SFB+AO (see the left panel of Figure~\ref{fig:4}) and \mbox{SFB + CCDK} (see the right panel of Figure \ref{fig:4}), the IST EoS is in very good agreement with the observational data. As in the non-superfluid case, the model predicts the surface temperature for the Cas A both with the fast cooling curve for $M_{DU}\simeq 2.0\, M_{\odot}$, as well as with the curves for low-mass stars with a heavy-elements envelope. For the SFB + CCDK gaps, the Cas A is only described with the curve for $M=1.8\, M_{\odot}$ also with a heavy-elements envelope.

\begin{figure}[H]
\centering
\includegraphics[scale=0.55]{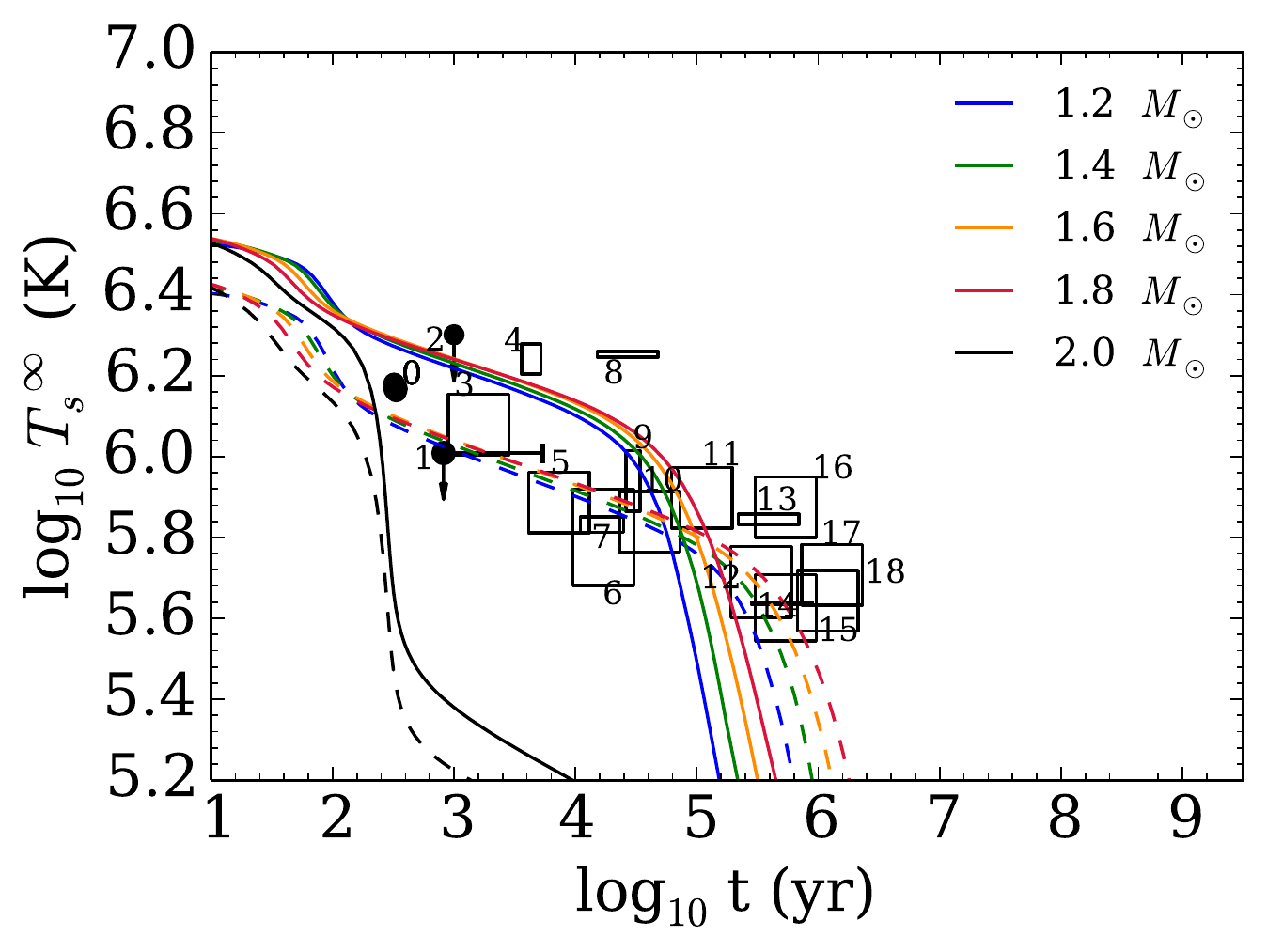}
\includegraphics[scale=0.55]{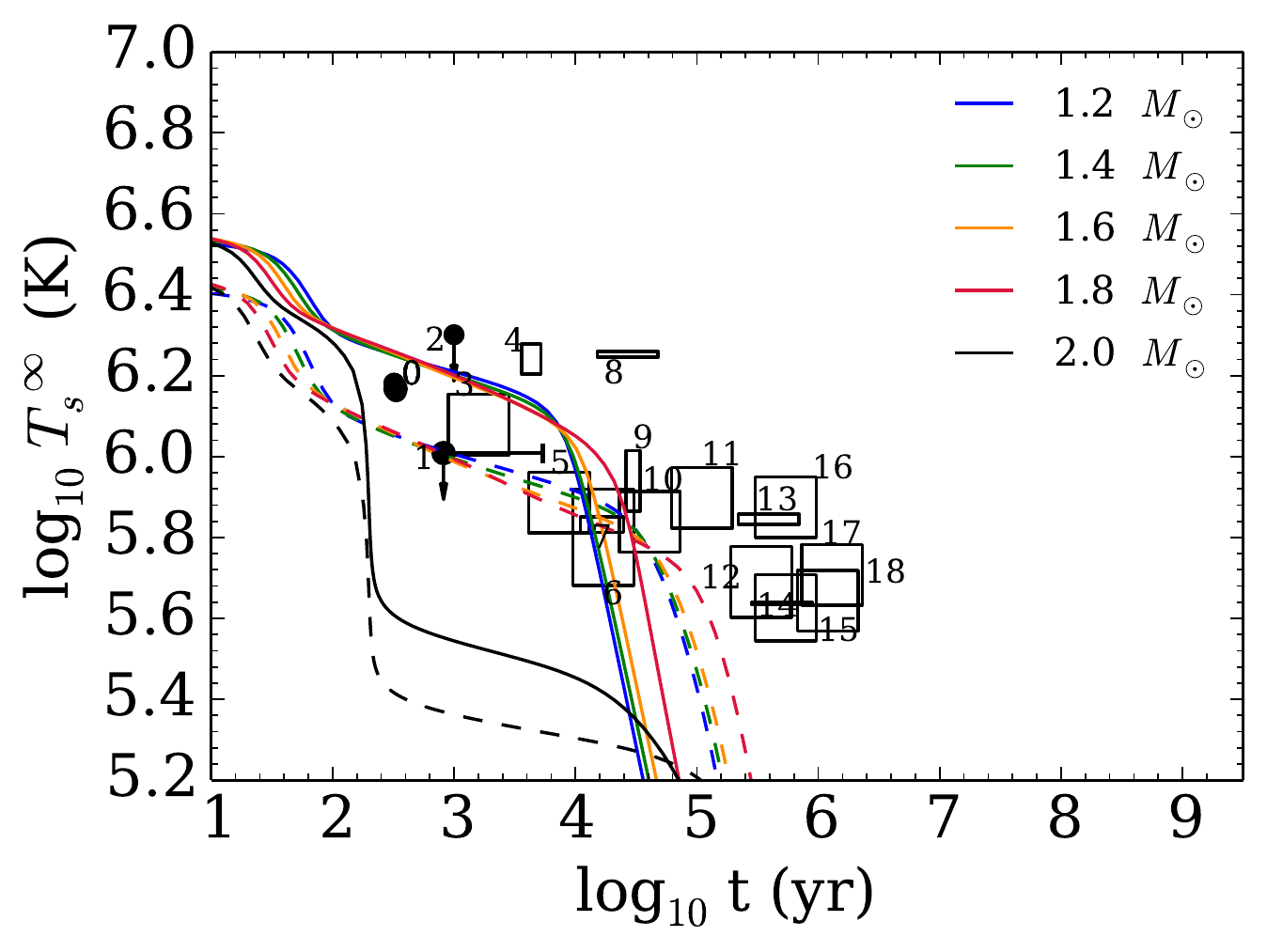}
\caption{The same as lower panel of Figure 4, but supplemented with the triplet neutron pairing in the core of the
star described by the T72 model \cite{Takatsuka1972} (left panel) and the AO model \cite{Amundsen1985b} (right panel). }                                                                                                                                                                                                                                                                                                                                                                                                                                                                                                                                                                                                                                                                                                                                                                                                                                                                                                                                                                                                                                                                                                                        
\label{fig:5} 	
\end{figure}

Finally, we studied the effect of the neutron triplet pairing on the cooling of NSs by considering a shallow gap with a maximum of the critical temperature at the saturation density $n_{0}$\mbox{ (T72 model \cite{Takatsuka1972})} and an extended gap with a maximum of $T_{c}$ at $2n_{0}$ (AO model \cite{Amundsen1985b}). As it shown in Figure \ref{fig:5}, the combination of models SFB (for $n^{1}S_{0}$)+T72 (for $n^{3}P_{2}$)+CCDK (for $p^{1}S_{0}$) (left panel) and SFB (for~$n^{1}S_{0}$)+AO (for $n^{3}P_{2}$)+CCDK (for $p^{1}S_{0}$) (right panel) give a qualitatively similar results. Adding the $n^{3}P_{2}$ pairing in the core of NSs leads to more rapid cooling and makes it incompatible with most of the observational data. Therefore, we conclude that, within our model, neutron pairing in the triplet channel is inconsistent with the observational data.

\subsection{Description of the Cas A Temperature Evolution}
\label{Cas}

\begin{figure}[H]
\centering
\includegraphics[scale=0.55]{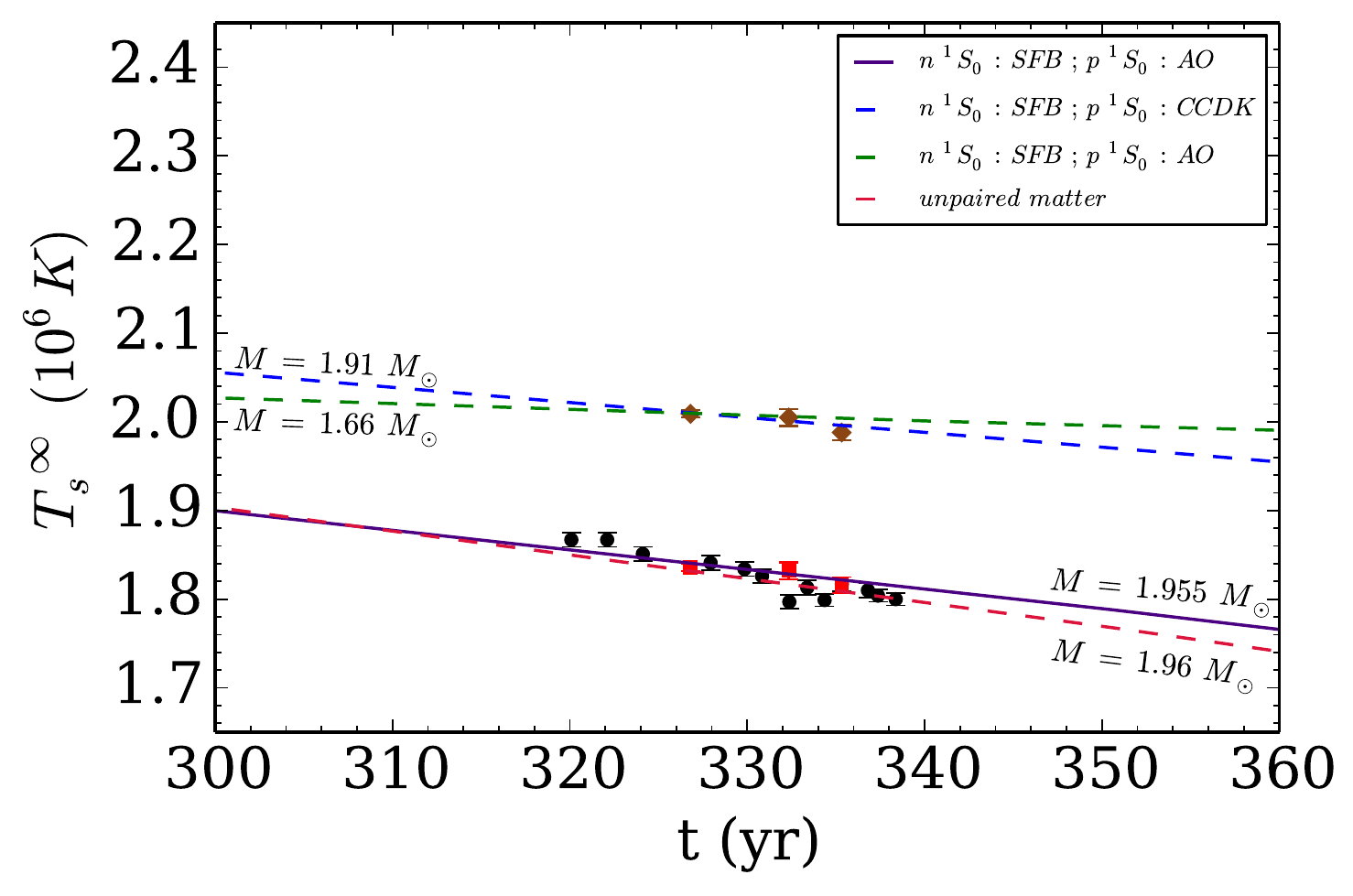}
\caption{Temperature decrease rate for the NS in Cas A. The brown diamonds ($N_{H}$ of 2006 \cite{2018ApJ...864..135P}) and red squares ($N_{H}$ of 2015 \cite{2019AIPC.2127b0007H}) correspond to Chandra ACIS subarray mode data. Black circles indicate temperature measured using Chandra ACIS-S Graded spectra \cite{2019MNRAS.484..974W}. Fit results are shown in blue, green, purple, and red.}                                                                                                                                                                                                                                                                                                                                                                                                                                                                                                                                                                                                                                                                                                                                                                                                                                                                                                                                                                                                                                                                                                                        
\label{fig:6} 	
\end{figure}

The fit results of the latest Cas A observational data within the IST EoS are presented on \mbox{Figure  \ref{fig:6}}. Namely, we find 
that the three data points (brown diamonds on \mbox{Figure \ref{fig:6}}) reported by Posselt\& Pavlov~\cite{2018ApJ...864..135P} can be described by a rapidly cooling $1.91\,M_{\odot}$ star with light-elements atmosphere (blue dashed curve). Despite a slightly slower cooling rate, the medium-mass $1.66~M_{\odot}$ star with light-elements envelope (green dashed curve) indicates a good agreement with the data. Another set of 13 Chandra ACIS-S Graded data points for constant interstellar absorption column \cite{2019MNRAS.484..974W} is depicted as black circles on Figure \ref{fig:6}. The red squares correspond to three points measured by Posselt\& Pavlov~\cite{2018ApJ...864..135P}, normalized by the observations of 2015 \cite{2019AIPC.2127b0007H}. As far as one can see, these three points together with Chandra ACIS-S Graded data can be simultaneously fitted with a fast cooling massive star of $1.96\,M_{\odot}$ with unpaired matter in its interior (red dashed curve), as well as with a  $1.955\,M_{\odot}$ star  with inclusion of neutron  and proton pairings in the singlet channel (purple solid curve). Both of these high-mass stars have switched-on DU processes that lead to such a fast temperature decrease. Therefore, we can conclude that the IST EoS is able to reproduce the apparent fast cooling of Cas A, even without the inclusion of neutron superfluidity and proton superconductivity. 

\section{Conclusions}
\label{Concl}

We studied the cooling of NSs with the novel IST EoS that was previously applied to the analysis of nuclear matter properties, heavy-ion collision experimental data, and was recently generalized for the description of matter inside NSs. The considered model parameterisation is in good agreement with the GW170817 and the recent measurements of the most massive NSs. For the stars with $M \geq 1.91\, M_{\odot}$ the model allows for the occurrence of DU processes that leads to much faster cooling. An effect of neutron pairing in $^{1}S_{0}$ and $^{3}P_{2}$ channels, as well as proton pairing in $^{1}S_{0}$ channel, is thoroughly analysed with different superfluidity/superconductivity models. 
In addition, the effects of different types of envelopes on the thermal evolution of NSs were also compared. We show that, for unpaired matter, the obtained cooling curves are fully consistent with the currently existing observational data.

We report that an explanation of the rapid temperature drop of Cas A does not necessarily require an inclusion of neutron superfluidity and/or proton conductivity in the model. The surface temperature of Cas A reported in Refs. \cite{2019MNRAS.484..974W, 2019AIPC.2127b0007H} is reproduced, within the IST EoS, by the cooling curves for medium and high-mass stars. Thus, a medium-mass $1.66~M_{\odot}$ NS with an envelope of light elements offers a reasonably good description and, at the same time, is the most congruous with the observational results. Alternatively, the data can be explained by the unpaired matter with a rapidly cooling $1.96~M_{\odot}$ star, as well as with a $1.955~M_{\odot}$ star (light-elements envelope) with included proton and neutron pairing in the singlet channel. Another set of measurements of the surface temperature of Cas A \cite{2018ApJ...864..135P} can be explained by a heavy-mass $1.91~M_{\odot}$ star with an allowed DU process.

The two considered models of the triplet neutron pairing in the core of the star (a shallow and an extended neutron superfluidity) lead to too rapid cooling of all NSs, such that the old stars \mbox{($t > 10^{5} $ yr)} cannot be reproduced. Thus, we can conclude that the calculations favour the vanishing neutron triplet pairing gap. 

As a future step, we plan to extend the list of particles by including heavy baryons and mesons. Despite their small effect on EoS, these particles do modify the thermal evolution of NSs and, therefore, have to be considered in the cooling simulations. 

\vspace{6pt}
\authorcontributions{Both authors contributed equally to this work. Both authors have read and agreed to the published version of the manuscript.}

\funding{V.S. acknowledges the support from the Funda\c c\~ao para a Ci\^encia e Tecnologia (FCT) within the project UID/04564/2020.}

\acknowledgments{We are thankful for the fruitful discussions and suggestions to M. Fortin, O. Ivanytskyi and C. Provid{\^e}ncia.}

\conflictsofinterest{The authors declare no conflict of interest.} 

\reftitle{References}

\end{document}